\documentclass[twocolumn]{aastex61}
\pdfoutput=1 
\usepackage{amsmath,amstext}
\usepackage[T1]{fontenc}
\usepackage{apjfonts} 
\usepackage{booktabs}
\usepackage[figure,figure*]{hypcap}

\newcommand{\makeinvisible}[1]{}

\shorttitle{GW Sky Localization}
\shortauthors{Pankow et al.}

\begin{document}

\newcommand{\nuaffil}{Center for Interdisciplinary Exploration and Research in Astrophysics (CIERA)
and
Department of Physics and Astronomy,
Northwestern University,
2145 Sheridan Road,
Evanston, IL 60208,
USA}
\newcommand{\dsfpaffil}{LSSTC Data Science Fellow}
\newcommand{\cardiffaffil}{School of Physics \& Astronomy, Cardiff University, The Parade, Cardiff, Wales, UK, CF24 3AA}

\title{Improvements in Gravitational-Wave Sky Localization with Expanded Networks of Interferometers}

\author{Chris Pankow}
\affiliation{\nuaffil}

\author{Eve A. Chase}
\affiliation{\nuaffil}
\affiliation{\dsfpaffil}

\author{Scott Coughlin}
\affiliation{\nuaffil}
\affiliation{\cardiffaffil}

\author{Michael Zevin}
\affiliation{\nuaffil}

\author{Vassiliki Kalogera}
\affiliation{\nuaffil}

\begin{abstract}
A milestone of multi-messenger astronomy has been achieved with the detection of gravitational waves from a binary neutron star merger accompanied by observations of several associated electromagnetic counterparts. Joint observations can reveal details of the engines that drive the electromagnetic and gravitational-wave emission. However, locating and identifying an electromagnetic counterpart to a gravitational-wave event is heavily reliant on localization of the source through gravitational-wave information. We explore the sky localization of a simulated set of neutron star mergers as the worldwide network of gravitational-wave detectors evolves through the next decade, performing the first such study for neutron star -- black hole binary sources. Currently, three detectors are observing with additional detectors in Japan and India expected to become operational in the coming years. With three detectors, we recover a median neutron star -- black hole binary sky localization of 60 deg$^2$ at the 90\% credible level. As all five detectors become operational, sources can be localized to a median of 11 deg$^2$ on the sky. 

\end{abstract}

\section{Introduction}

The joint detection of gravitational waves from a binary neutron-star merger~\citep{2017PhRvL.119p1101A} and subsequent electromagnetic observations of an associated short gamma-ray burst~\citep{2017ApJ...848L..13A} and kilonova~\citep{2017ApJ...850L..39A} have added gravitational-wave observations to multi-messenger astronomy~\citep{2017ApJ...848L..12A}. The detection of events with electromagnetic, neutrino, and gravitational-wave detections can provide cross-cutting measurements of phenomena that would be otherwise inaccessible individually, for instance, an independent measurement of the Hubble constant via electromagnetic and gravitational-wave distance estimates~\cite{LVCHubble2017}. Additional multi-messenger events are awaited, as the advanced LIGO and Virgo detectors reach design sensitivity~\citep{2015CQGra..32b4001A, AdvLIGO}. In the era of multi-messenger astronomy, improved sky localization of gravitational-wave events will promote the routine and rapid identification of electromagnetic counterparts.
The study presented here focuses on compact binaries with at least one neutron star component, with a specific emphasis on neutron star -- black hole binaries. Both categories provide a wide variety of electromagnetic signatures~\cite{0004-637X-746-1-48}, such as gamma-ray bursts, afterglows, and kilonova. However, certain configurations of neutron star -- black hole binaries are unfavorable to such emission as the neutron star is engulfed by the black hole before disruption~\cite{2041-8205-791-1-L7}.

Localization on the sky with networks of gravitational-wave interferometers relies on relative signal arrival times and the realized signal amplitude in each detector. For example, a network of two instruments with perfect timing reconstruction would produce rings on the sky corresponding to locations with similar arrival times. Using consistency between the amplitude responses in widely-separated instruments can refine this result to a few disconnected arcs along the ring~\citep{2014ApJ...795..105S}. Furthermore, the inclusion of physical parameters of the source (intrinsic parameters such as spin) can help break degeneracies and further constrain sky localization~\citep{2008ApJ...688L..61V, 0264-9381-26-11-114007}: component spins misaligned with the orbital angular momentum can introduce ``knotting'' in the arc. However, with only two detectors, electromagnetic facilities would likely need to search hundreds to thousands of square degrees on the sky to adequately survey the full posterior probability distribution of sky localization.

The Virgo detector in Cascina, Italy joined the two US-based LIGO detectors for the last month of LIGO's second observing run. The addition of a third detector can substantially improve localization by introducing another set of baselines, and consequently, several intersecting rings on the sky of constant arrival time. The intersections of the rings identify source locations consistent with measured arrival time differences. Additionally, despite Virgo's lower sensitivity during O2 compared to the two LIGO detectors, the inclusion of data from Virgo provides another set of baselines as well as another point of comparison for the expected signal amplitude on arrival at separated instruments. This is exemplified by the LIGO-Virgo detections GW170814 and GW170817. GW170814~\citep{2017PhRvL.119n1101A}, the first three-detector observation of gravitational waves from a compact binary merger, was localized to 1160 deg$^2$ at the 90\% credible level with only the two LIGO detectors. The inclusion of Virgo reduced this area to only 100 deg$^2$. For the binary neutron star merger GW170817, the addition of data from Virgo reduced the LIGO-only 100 deg$^2$ sky localization to roughly 30 deg$^2$~\citep{2017PhRvL.119p1101A}, enabling the swift identification of the electromagnetic transients and host galaxy~\citep{2017ApJ...848L..12A}. Over the next decade, it is anticipated that sky localization of compact binary systems detected through gravitational waves will be confined to areas of a few tens to hundreds of square degrees~\citep{2016LRR....19....1A}, as all detectors reach design sensitivity.

Two additional ground-based interferometers are in the construction or planning stages. Kagra, a Japanese-built, cryogenically-cooled, 3-km interferometer in the Kamioka mine~\citep{2013PhRvD..88d3007A} is in the construction stage. Kagra is projected to begin science operations during or soon after 2018. In addition, construction is expected to begin on the LIGO-India detector in the early 2020s with possible operations beginning in 2024 \citep{ligoindia}. If all five instruments are active and equally sensitive, it is expected that sky localization regions for binary neutron star (BNS) systems will reduce to $\sim 1 - 10$ deg$^2$~\citep{2010PhRvD..81h2001W}.

In this study, we examine sky localization mainly for a set of neutron star -- black hole (NSBH) mergers detectable by the LIGO-Virgo network (HLV). We then study the same events in the LIGO-Kagra-Virgo (HKLV) and LIGO-India-Kagra-Virgo (HIKLV) detector configurations, following the expected progression of additions to the detector network. In general, the progression from three-, to four-, to five-instrument networks reduces the median integral sky area at the 90\% credible interval by about a factor of two between each detector configuration for a canonical population of NSBH sources. While multiple studies have examined sky localization with BNS systems~\citep{2011ApJ...739...99N,2012PhRvD..85j4045V,2014ApJ...795..105S,2014ApJ...784..119R,2015ApJ...804..114B,2016ApJ...825..116F}, this is the first study to examine the localization of a population of NSBH events with plausible component masses and spins in the advanced gravitational-wave detector era. We also perform a similar study for a smaller selection of BNS sources, as a comparison metric with past studies.

\section{Gravitational-wave Sky Localization}
\label{sec:sky_loc}

The amplitude, time of arrival, and phase of the gravitational-wave strain at a given instrument is dependent on the source location. The measured strain has an intrinsic source-dependent amplitude, and two factors (the so-called ``antenna patterns") determining the geometric sensitivity to either polarization of the gravitational wave. A single interferometer, and hence only a single measurement of the combination of two amplitudes, is unable to independently measure responses to both gravitational-wave polarizations, nor a time-of-arrival difference, and is therefore unsuitable for source sky localization.
However, localization is possible with two or more interferometers, as each additional detector provides an independent measure of the difference in time of arrival along the sites' baseline. Each difference corresponds to a ring of locations in the sky, with the center of the ring formed by the baseline projected into the celestial sphere.  The intersections of those rings for each detector pair indicate most probable source locations on the sky. As the number of detectors increases, specific intersections are favored, further reducing sky localization regions. This is equivalent to the technique of triangulation~\citep{2011NJPh...13f9602F}. Rings of constant arrival time have widths proportional to source timing uncertainty. In addition to the time of arrival precision (and hence timing uncertainty), the precision with which phase of arrival is determined is also crucial to the localization size~\citep{2016PhRvD..93b4013S}.

The orientation of the gravitational-wave detector relative to the incident source direction further affects localization. Different detectors are not equally sensitive to the same source location, as gravitational-wave interferometers are not omni-directional and do not have aligned zenith axes. As additional detectors are constructed, the constraints on which relative amplitudes (corresponding to antenna-pattern values) are supported by the data become tighter, since several disparate amplitudes must be matched self-consistently by the signal projected at each site. Consistency in the expected detected amplitude of an incoming wave between different sites can also further reduce localization regions. In general, having differing amplitude response from geographically-separated sites provides better localization, as was notable for GW170817~\citep{2017PhRvL.119p1101A}. For a given network configuration, it is possible to extract the expected error region from the Fisher information matrix as a timing-weighted sum over the angles formed by the normal of the planes of each non-degenerate instrument triple and the wave propagation direction~\citep{2010PhRvD..81h2001W}. Geographically, the placement of new instruments will have varying effects on localization, depending on location and orientation relative to the remainder of the network. Previous studies have examined the optimal placement of new instruments~\citep{2006PhRvD..73l4014S,2015CQGra..32j5010H}.

Localization also depends on source parameters, with BNS systems generally achieving better localization than NSBH systems. This is due to the frequency content of the signal, measured by its effective bandwidth. Effective bandwidth is the second noise-weighted moment in frequency of a source waveform (see Equation 2 of \citealt{2011CQGra..28j5021F}). Thus, sky localization for sources with equal signal-to-noise ratio (SNR) will change with differing effective bandwidth. NSBH systems typically have smaller effective bandwidths than BNS systems, due to more compact frequency content. Based on effective bandwidth alone, a canonical NSBH system (1.4 $M_{\odot}$ + 10 $M_{\odot}$) is expected to have 1.3$^2$ times larger localization error regions than their canonical BNS counterparts (1.4 $M_{\odot}$ + 1.4 $M_{\odot}$), at equal SNR. However, effective bandwidth and SNR in isolation are not sufficient to fully describe the sky localization capabilities of a given network.

Other source parameters, such as spin, can further affect localization through the effective bandwidth. Large spin magnitudes, when their direction is orthogonal to the plane of the binary, shift the frequency content of the waveform relative to the non-spinning case~\citep{rotbh,2009PhRvD..80l4026R}. Thus the effective bandwidth is changed because this power now resides in a frequency region with different spectral sensitivity, changing the expected size of localization regions. The components of the spin projected into the orbital plane also affect localization~\citep{2008ApJ...688L..61V,0264-9381-26-11-114007}, but nominally through breaking degeneracies in the parameter space, rather than through the effective bandwidth. 

\section{Posterior Sampling and Event Population}

A fixed set of simulated gravitational-wave events is analyzed in three different detector configurations: HLV, HKLV, and HIKLV. All event properties are left unchanged between each network configurations. We employ \texttt{lalinference\_mcmc}~\citep{2015PhRvD..91d2003V}, one of the standard Markov-Chain Monte Carlo samplers developed within the LIGO-Virgo collaborations to produce Bayesian posteriors of binary waveform properties. To ensure that the effect of correlation with other source properties is accounted for, we sample all source properties and orientation parameters. However, only the sky area is examined in detail. We refer the reader to \cite{2017arXiv170308988G} for how accurately other properties such as masses and spins would be measured with additional detectors. As with previous studies, our simulations are performed noise-free to separate the localization uncertainties from effects arising from realizations of the noise~\citep{2014ApJ...784..119R,2015ApJ...804..114B}. All instruments are assumed to have identical advanced LIGO design sensitivity curves~\citep{2015CQGra..32g4001L} with the exception of Virgo which assumes its own advanced Virgo curve~\citep{2015CQGra..32b4001A,2016JCAP...11..056P}.

\subsection{Neutron Star -- Black Hole Binaries}

We make use of a population of NSBHs, with waveforms modeled using the \texttt{IMRPhenomPv2}~\citep{2014PhRvL.113o1101H} family. 
Previous studies addressing NSBH have assumed fiducial, fixed masses and neglect spin. Known X-ray binaries (XRB), one of the most likely evolutionary paths of NSBH, have a handful of mass and spin measurements~\cite{0004-637X-741-2-103,0004-637X-800-1-17,2016A&A...587A..61C}. While a detailed population analysis is beyond the scope of this work, the values measured span the range explored here. Therefore, in the absence of well measured population statistics, we note that uniform distributions are plausible and have the additional advantage of mapping reasonably well on to the priors adopted by gravitational-wave parameter estimation~\cite{2015PhRvD..91d2003V}.
Masses are drawn from a uniform distribution between 5 -- 30 $M_{\odot}$ for BH and 1 -- 3 $M_{\odot}$ for NS. The black hole spin is distributed isotropically with dimensionless magnitudes ($a=|S/m^2|$) uniform up to 0.99. The population is selected to represent a plausible set of detections with the three-instrument HLV network, described in \cite{2017ApJ...834..154P}. The set of posteriors from \cite{2017ApJ...834..154P} are used for the HLV network studies, without modification\footnote{In \cite{2017ApJ...834..154P}, five configurations of fixed parameter sets were examined. In this work, we use the posteriors with no fixed source parameters.}.

\subsection{Binary Neutron Stars}

When comparing the effect of source properties, BNS populations should have less variation in sky localization: their mass ranges are narrower --- thus leading to less variation in effective bandwidth --- and their spins are not expected to be high enough to induce measurable precessional effects~\citep{2016ApJ...832L..21A}. It is unlikely that other properties, such as tidal deformability are significant enough to affect sky localization. As BNS localization is addressed in several previous studies, we restrict our study to only three representative BNS systems, each with different spin magnitudes but consistent masses (1.4 $M_{\odot}$ + 1.4 $M_{\odot}$). 
We examine three different spins: zero-spin, component spin magnitudes equal to the largest spin measured in a binary of NS with dimensionless spin magnitude $a=0.05$~\citep{2009CQGra..26g3001K}, and spin magnitudes near the fastest observed neutron star spin at $a=0.4$~\citep{2008ApJ...675..670F}.
In all cases, the spin direction is oriented in the plane of the binary to capture the range of variation induced in the waveform by the precessing spins. Each system is then scaled to distances corresponding to SNR of 12, 14, 16, and 20 to simulate various potential detection ranges over each network configuration.

\section{Results} \label{sec:res}

The following sections present the resultant credible regions for the transition between HLV to HKLV and HKLV to HIKLV. It is important to recall that this is equivalent to viewing the \emph{same} population of events, but with additional instruments added to the network. In general, this means that the SNR of each event receives contributions from additional instruments. The overall improvement in sky localization is thus split between both the increase in SNR (which improves the timing uncertainty measurement in the network) as well as the additional constraints added by amplitude and phase consistency over the network.

When visualizing distributions which would nominally be shifted across the sky by the Earth's rotation, we instead use Earth-fixed coordinates. This is achieved by aligning the zero of the right ascension with the prime meridian to represent sky alignment with the Earth at the event time. This representation highlights any correlations between sky localization size and network orientation which is necessarily fixed to the frame of the Earth.

\subsection{Neutron Star -- Black Hole Binaries} \label{sec:res_nsbh}

\begin{figure*}[htbp]
\includegraphics[trim={1cm 2cm 1cm 2cm },clip,width=\columnwidth]{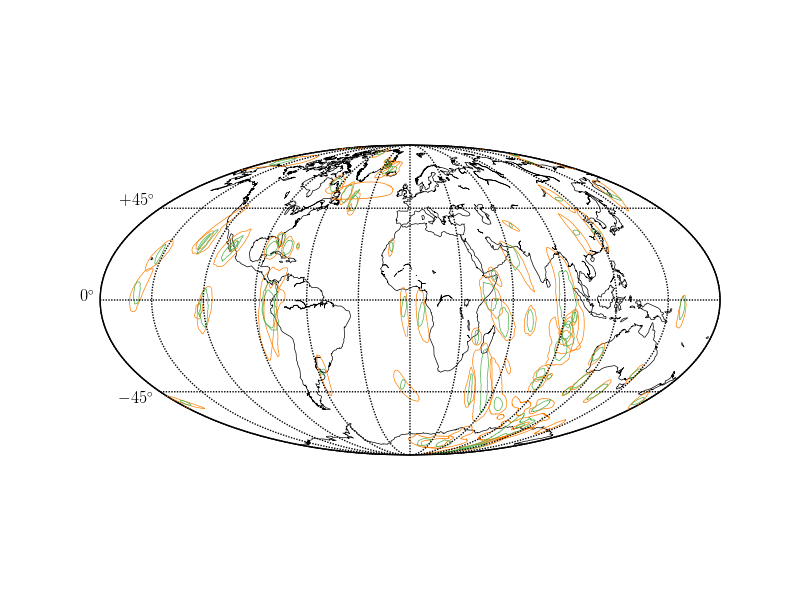}
\includegraphics[trim={1cm 2cm 1cm 2cm },clip,width=\columnwidth]{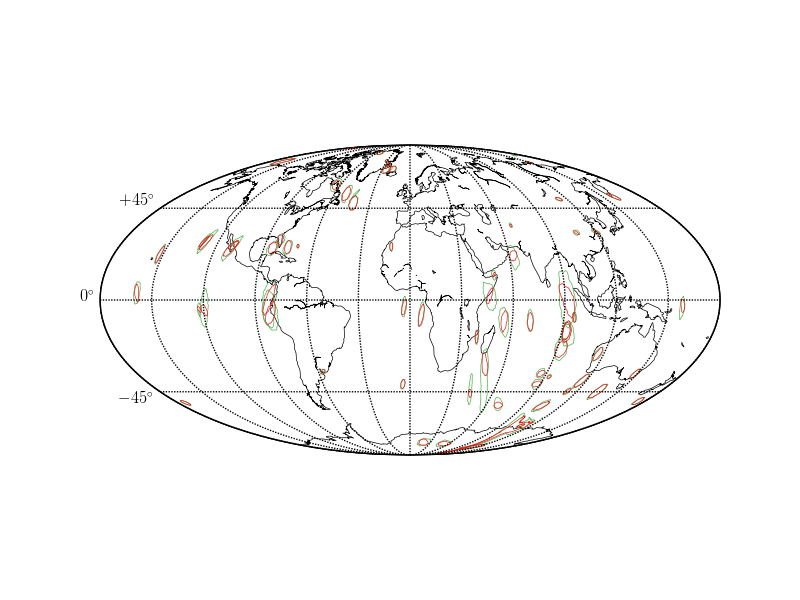}
\caption{\label{fig:3to4_skymap} \textit{Left:} 90\% credible regions for NSBH sky localization with the HLV (orange) and HKLV (green) networks plotted on the sky in an Earth-fixed geographic coordinate frame. \textit{Right:} Same, but for the HKLV (green) to HIKLV (red) transition.}
\end{figure*}

The panels in Figure \ref{fig:3to4_skymap} show the transition in sky localization capability for the same set of NSBH events realized in each detector configuration. Specifically, the left panel in Figure \ref{fig:3to4_skymap} shows the same events in the transition from LIGO-Virgo to LIGO-Kagra-Virgo. The improvement in sky localization is universal.

The right panel of Figure \ref{fig:3to4_skymap} shows a similar transition as the LIGO-India detector is activated. Here, the decrease is less noticeable as arc-like regions have already been reduced to more symmetrical ellipses. The best improvement is now apparent over the Indian Ocean, northeast of Australia, as well as a continued decrease in error region size over the equator in the Pacific. This effect is due to the geographic location of LIGO-India, as these events would be nearly directly overhead and thus well oriented for detection.

\begin{figure*}[htbp]
\includegraphics[clip,width=\columnwidth]{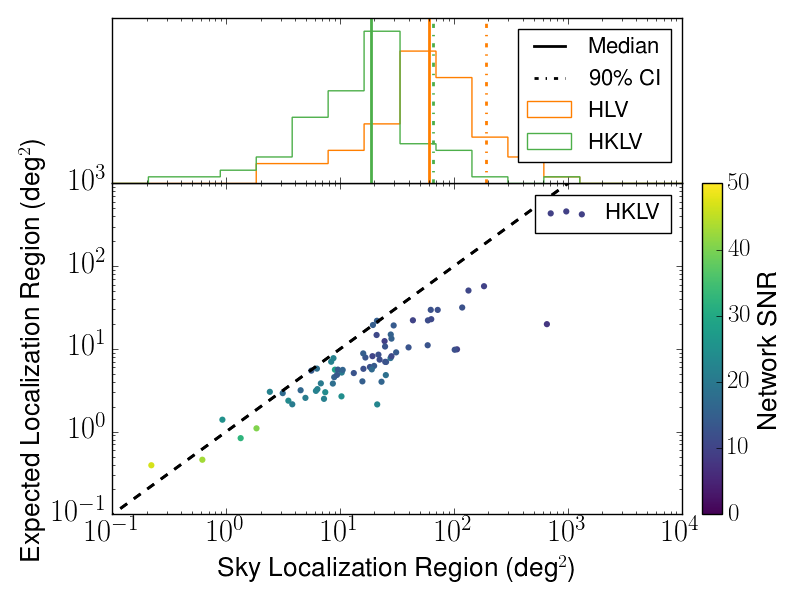}
\includegraphics[clip,width=\columnwidth]{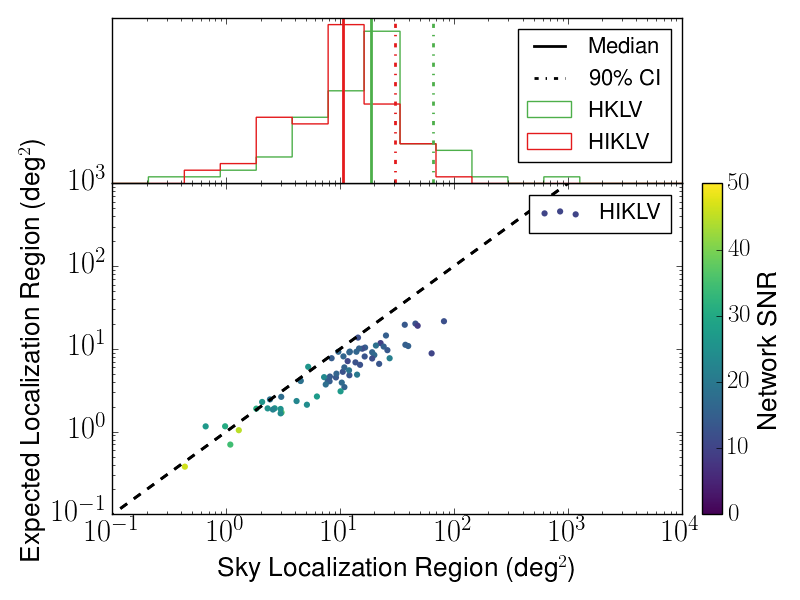}
\caption{\label{fig:scatter} \textit{Left:} scatterplot and histogram of 90\% credible regions for NSBH sky localization. The histogram shows the distribution of sky localizations for both HLV (orange) and HKLV (green) detector networks. The scatterplot displays localization of all NSBH events in the HKLV network, along with the expected sky localization according to Equation 41 in \cite{2010PhRvD..81h2001W}. Events are colored by their network SNR. For comparison purposes, a dashed line is displayed for equal expected sky localization error region and computed sky localization error region. \textit{Right:} Same, but for the HKLV (green) to HIKLV (red) transition. The scatterplot now contains events in the HIKLV network.}
\end{figure*}

We compare our NSBH localizations to the expected sky localizations according to Equation 41 in \cite{2010PhRvD..81h2001W} in Figure~\ref{fig:scatter}, finding generally consistent results. The histograms indicate the median and 90\% credible intervals for each network configuration of these events, which are also recorded in Table~\ref{tbl:nsbh_cr}.
The notable scatter between expectation and reality may arise from systematic underestimation, particularly at low SNR where assumptions inherent to the analytic approximations break down~\cite{PhysRevD.77.042001}.
 Overall, the median and 90\% intervals decrease by a factor of about three between the three- and four-detector transitions and a factor of two as a fifth detector is added. We also explore the relationship between network SNR and sky localization for HKLV and HIKLV configurations in Figure~\ref{fig:scatter}, recovering the expected relationship as suggested in~\cite{2015ApJ...804..114B}.

\begin{table}[htbp]
\centering
\begin{tabular}{cccc}
	\cmidrule[\heavyrulewidth](lr){2-4}
    & HLV & HKLV & HIKLV \\
	\midrule
	50\% & 60 & 19 & 11 \\
	90\% & 193 & 66 & 30 \\
    \bottomrule
\end{tabular}
\caption{\label{tbl:nsbh_cr} Summary of 50\% (median) and 90\% fractions of the 90\% credible regions for the NSBH event study. Values are represented in units of square degrees.}
\end{table}

\subsection{Binary Neutron Stars} \label{sec:res_bns}

\begin{figure}[htbp]
\includegraphics[width=\columnwidth]{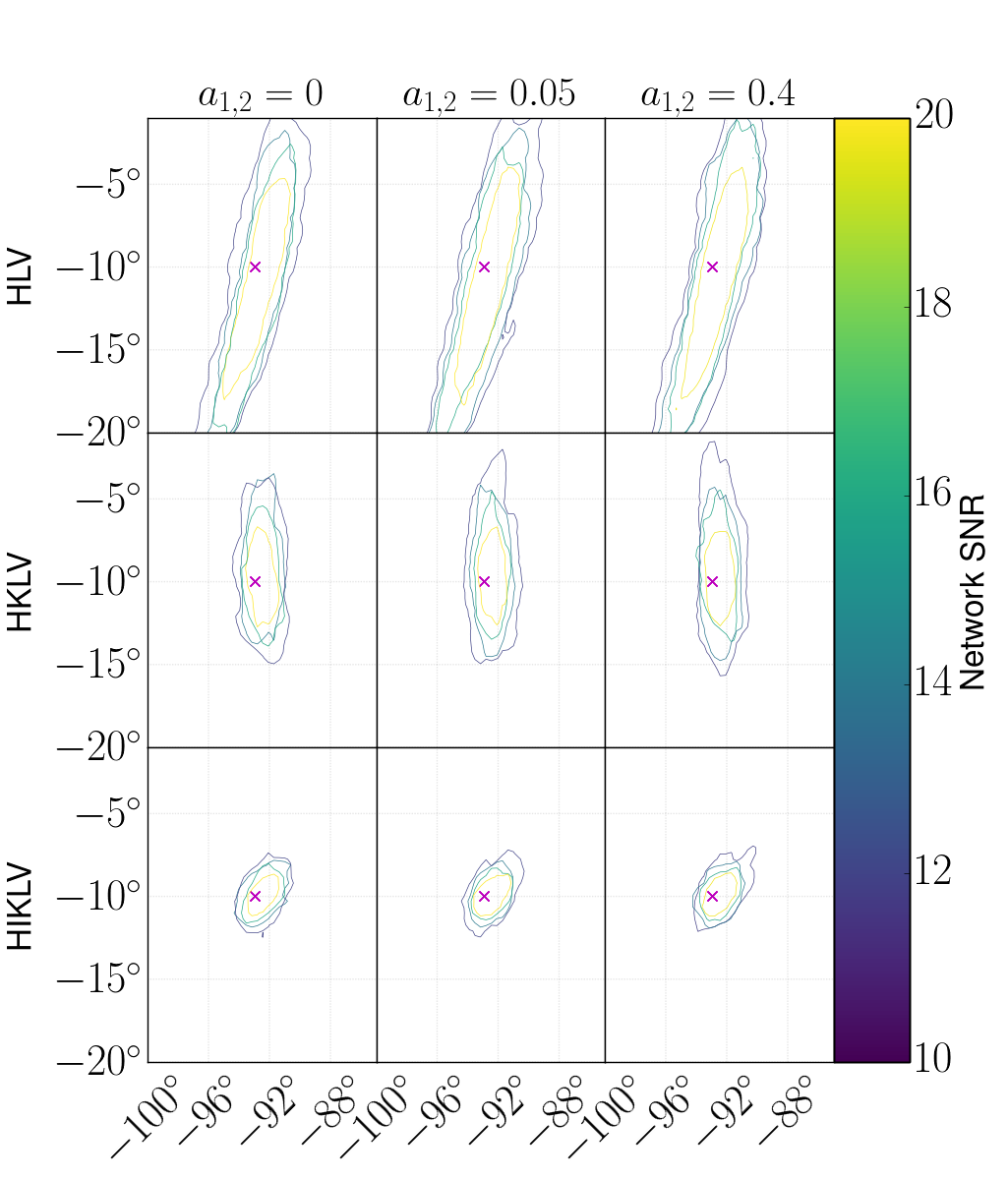}
\caption{\label{fig:bns_skymap} The 90\% localization regions for a canonical BNS system (1.4 $M_{\odot}$ + 1.4 $M_{\odot}$) at each of the three network configurations and spin magnitude values. Superimposed in each panel are the localization regions at different network SNRs (12, 14, 16, and 20). Networks HLV, HKLV, and HIKLV are read top to bottom, and spins start on the left, iterating through 0, 0.05, and 0.4. Spin configurations examined here have little influence on the region size and shape.}
\end{figure}

The resulting localization regions for the full parameter set BNS study are displayed in Figure \ref{fig:bns_skymap} for all considered spin configurations. We recover the expected scaling of sky localization areas with SNR. In general, there is no discernible difference between signals with zero and non-zero spin when considered at the same SNR and network configuration. While only one sky location has been examined, we expect similar results for most BNS signal parameters, scaled appropriate by fiducial sky location. Table \ref{tbl:bns_err_reg} contains a full list of localization areas at 90\% credibility.

\citet{2011CQGra..28j5021F} predicts a localization region size between $10$ and $150$ deg$^2$ for a canonical BNS event in the HLV network with a reference SNR near 14. We obtain a value of 43 deg$^2$, near the median of areas predicted in \citet{2011CQGra..28j5021F}, for our comparable zero-spin and SNR of 14 localizations in the HLV configuration. As Kagra is added to the network\footnote{Note that \citet{2011CQGra..28j5021F} and most other previous studies refer to Kagra as `J'.}, \citet{2011CQGra..28j5021F} predicts localization regions to shrink to between $4$ and $22$ deg$^2$. We recover a localization region around $21$ deg$^2$ for the HKLV network, remaining broadly consistent with the predictions of \citet{2011CQGra..28j5021F}. While the inclusion of LIGO-India was not examined in \citet{2011CQGra..28j5021F}, we recover a median localization of 9.3 deg$^2$ for the HIKLV network.

\begin{table*}[t]
\centering
\begin{tabular}{c|ccc|ccc|ccc}
    \cmidrule[\heavyrulewidth]{2-10}
    \multicolumn{1}{c}{} & \multicolumn{3}{c}{HLV} & \multicolumn{3}{c}{HKLV} & \multicolumn{3}{c}{HIKLV} \\
	\cmidrule{2-10}
	SNR & 0 & 0.05 & 0.4 & 0 & 0.05 & 0.4 & 0 & 0.05 & 0.4 \\
	\midrule
	12 & 74 & 80 & 80 & 30 & 34 & 34 & 12 & 13 & 12 \\
	14 & 43 & 56 & 58 & 21 & 21 & 22 & 9.3 & 8.1 & 9.0 \\
	16 & 37 & 35 & 45 & 15 & 13 & 15 & 6.1 & 5.8 & 5.9 \\
	20 & 22 & 24 & 24 & 8.0 & 7.8 & 8.3 & 3.1 & 3.8 & 3.9 \\
    \bottomrule
\end{tabular}
\caption{\label{tbl:bns_err_reg} The 90\% localization region values for BNS runs in square degrees. Each set of three columns corresponds to a network configuration (top), and a component spin magnitude (below network). Each row corresponds to the same event with a distance adjust such that the SNR values are 12, 14, 16, and 20.}
\end{table*}

\subsection{Comparison With Previous Studies}

We can compare our result to several studies of advanced network sky localization which have already been performed. We compared our results to those from analytic expectations of \citet{2010PhRvD..81h2001W}  and \citet{2011CQGra..28j5021F} throughout Sections~\ref{sec:res_nsbh} and~\ref{sec:res_bns}, respectively. The expression used to calculate the expected error regions from \citet{2010PhRvD..81h2001W} produces values which are consistent with both \citet{2011CQGra..28j5021F} and our studies. In addition, our study remains broadly consistent with the analytic results of~\citet{2011PhRvD..84j4020V}, recovering similar scalings in localization improvement for successive detector networks for a canonical BNS to our results in Table~\ref{tbl:bns_err_reg}.

Studies such as \citet{2014ApJ...795..105S} have examined the sky localization capabilities of two- and three-interferometer networks of LIGO and LIGO-Virgo, using more realistic assumptions. In particular, they carried out an end-to-end study of the localizations from the outputs of one compact binary search pipeline used in the first two observing runs~\citep{PhysRevD.95.042001}. They use a stream of data with colored, but Gaussian, noise and apply a minimum network SNR cut of 12 for each event. \citet{2014ApJ...795..105S} projected that Virgo would join the LIGO instruments during the second observation run at the sensitivity indicated in~\citet{2016LRR....19....1A}. They examine a population of BNSs in the HLV configuration, obtaining a median sky localization of $235$ deg$^2$ at the 90\% credible level. For our study of NSBH localization, we obtain a median sky localization of $193$ deg$^2$, seemingly contrary to the expectation that BNS sources are better localized than NSBHs. However, the fiducial networks in our study are have better sensitivities over a given frequency range compared to \citet{2014ApJ...795..105S}, leading to improvements in the effective bandwidth measured by a given instrument. So, our improved localization from that predicted by \citet{2014ApJ...795..105S} can be attributed to our use of 2018-era instruments as opposed to the less-sensitive 2016 configuration used in \citet{2014ApJ...795..105S}.

\citet{2014ApJ...784..119R} resolve sky localization for a BNS population in which each event realizes a network SNR of 20. On the contrary, our NSBH study samples a wide distribution of network SNRs, meeting a minimum SNR threshold\footnote{We require an SNR of at least 5.5 in the second-loudest detector.}. Therefore, our study probes the SNR regime at which detections are most likely to occur, while~\citet{2014ApJ...784..119R} ignores this low network SNR range. Nevertheless, our Figure~\ref{fig:3to4_skymap} is qualitatively comparable to Figures 6 and 7 in \citet{2014ApJ...784..119R}. Particularly, the increased isotropy of a three- to four-instrument network and the subsequent, near-universal reduction of arc-like error regions to more compact forms are comparable.

\cite{2012PhRvD..85j4045V} examine the improvement between the HLV and HKLV configurations for a set of canonical (1.4 $M_{\odot}$ + 1.4 $M_{\odot}$) BNS events, distributed at a fixed distance but otherwise isotropically in orientation. \citet{2012PhRvD..85j4045V} obtain a median sky localization of 10 deg$^2$ at the 90\% credible level with a four-site network, while for a similar network we obtain a median of 19 deg$^2$ for our set of NSBHs. This increase in sky localization by a factor of 1.9 is consistent with the na\"ive expected improvement of $1.35^2 \approx 1.8$, based on the decrease of effective bandwidth as discussed in Section \ref{sec:sky_loc}.

\citet{2013ApJ...767..124N} track localization improvement for a set of 5 $M_{\odot}$ + 1.4 $M_{\odot}$ NSBH events, with non-spinning binary components. For a three-detector network, \citet{2013ApJ...767..124N} recover a median localization of 50 deg$^2$ with a maximum localization of 170 deg$^2$ for all events studied. \citet{2013ApJ...767..124N} recover an analogous five-detector median localization of 6 deg$^2$ with a maximum of 65 deg$^2$. These results are consistent with our NSBH localizations recorded in Table~\ref{tbl:nsbh_cr}, with slight differences largely attributed to differences in NSBH mass and spin.

\citet{2016arXiv161201471C} examine a population of BNS in both a three instrument HLV as well as five instrument HIKLV configuration with a volumetrically and isotropically distributed population. Our fiducial event in the HLV configuration falls above their median for all SNRs, but is comparable for their median in the five instrument configuration, when the SNR 20 case is considered.

\section{Electromagnetic Follow-Up}

We can use the electromagnetic follow-up campaigns of GW150914~\citep{2016ApJ...826L..13A} and GW170817~\citep{2017ApJ...848L..12A} as examples of how the electromagnetic observation of a potential NSBH event would develop. The follow-up of GW150914 and GW170817 included several facilities which are likely to continue operation as the three-instrument network reaches design sensitivity. 
Potential electromagnetic counterparts to BNS and NSBH signals should be accessible to many of these instruments, as the peak gravitational-wave emission exceeds the instruments' limiting sensitivities for both BNS 
(no further than $\sim 500$ Mpc, $z \sim 0.1$) and a canonical NSBH (approximately $\sim 2$ Gpc, $z \sim 0.4$ for a 10 $M_{\odot}$ + 1.4 $M_{\odot}$ system)~\citep{2017arXiv170908079C}. If we are able to confidently separate the true source from a transient background, then the limiting factor for electromagnetic counterpart identification is slew and observation time versus field of view (FOV), under ideal observing conditions.

GW170817 was relatively nearby and well-localized, allowing a manageable area to be searched through electromagnetic campaigns. GW170817's localization lies well below the median expected localization for the second observing run~\citep{2016LRR....19....1A} and near the median expected for the fully advanced LIGO-Virgo network~\citep{2012PhRvD..85j4045V}. Given the population of NSBHs in this work, it is likely that a NSBH source will lie further away than GW170817 (given the accessible volume of space for just the three-instrument network) and thus the sky localization will expand accordingly. If the cadence and FOV of current and upcoming telescopes are typical of follow-up instruments for NSBH events, prospects of joint detection before a fourth instrument activates will be challenging. Several of the current and upcoming wide field instruments, such as the Dark Energy Camera~\citep{Flaugher:2015pxc}, PanSTARRS~\citep{2017Natur.551...75S}, ZTF~\citep{2016PASP..128h4501B}, and LSST~\citep{2008arXiv0805.2366I}, have FOVs between 1 deg$^2$ and 10 deg$^2$ while slew and observation times can vary significantly (see Table 4 in~\cite{0004-637X-814-1-25}). In the four- and five-detector era, NSBH sources will typically be localized to less than 20 deg$^2$, thus allowing these instruments to survey localization regions with a few to several pointings. Identification of an electromagnetic counterpart in these localization regions depends on several factors including magnitude and observation time. Such factors will motivate optimal follow-up campaigns to search localization regions for possible multi-messenger sources.

\section{Conclusion}

We have made the first examination of the localization capabilities for a set of astrophysically-motivated NSBH events projected through the expected evolution of the advanced gravitational-wave network. When additional independent baselines are added, the coverage of the network becomes more isotropic and reduces 90\% confidence regions drastically, sometimes by an order of magnitude between the three- and five-instrument networks. With the five-instrument network, there are few patches of the sky where sky localization would exceed $100$ deg$^2$, and in some cases, the required search area is less than a few square degrees. NSBH sources are localized at the 90\% credible region to a median of 60 deg$^2$ in the three-detector era. Analogous localization is reduced to 19 deg$^2$ in the four-detector era and 11 deg$^2$ with five detectors. 
Detected event populations tend to "pile up" in regions of the sky where the network response is most sensitive, and the antenna response of the HLV network is qualitatively different from HKLV and HIKLV.  Adding an instrument can never decrease the localization capability, and broadly speaking, adding additional long baselines (specifically those added with Japan and India) provides the potential for better localization. However, localization capability is not necessarily correlated with network response --- and hence where in the sky events are detected --- so it does not immediately follow that the medians presented here are conservative.
These improvements should facilitate complete and regular electromagnetic follow-up campaigns by many observatories at nominal coverage rates. The results presented here are in broad agreement with both analytical treatments as well as previous studies which have addressed the question of BNS localization in late advanced detector eras and a variety of potential network configurations.

Finally, while we have pinned our studies to the fully advanced LIGO design sensitivity, it is likely that this will not be representative of the state of GW interferometry in the mid 2020s. Particularly, it is possible that the five instrument network may have at least some of the interferometers in a so-called `A+' configuration which may add another factor of two or three to the ranges from the advanced LIGO/Virgo design sensitivity~\citep{2017CQGra..34d4001A}. If realized, these instruments will further improve sky localization capabilities with both increased SNR and narrower timing uncertainty.

\acknowledgments
CP and VK are supported by the NSF grant PHY-1607709, parts of this work were performed at the Aspen Center for Physics, which is supported by National Science Foundation grant PHY-1607761.
CP also acknowledges support by the Center for Interdisciplinary Exploration and Research in Astrophysics (CIERA).
EAC thanks the LSSTC Data Science Fellowship Program; her time as a fellow has benefited this work. 
SC is supported by the National Science Foundation, award
INSPIRE 15-47880. 
MZ acknowledges support from the Reach for the Stars GK-12 program, NSF grant DGE-1007911.
EAC and MZ also acknowledge support from the Illinois Space Grant Consortium Graduate Fellowship Program.
This research was supported in part through the computational resources from the Grail computing cluster --- funded through NSF Gravitational Physics --- and staff contributions provided for the Quest high performance computing facility at Northwestern University which is jointly supported by the Office of the Provost, the Office for Research, and Northwestern University Information Technology.
Additionally, the authors would like to thank Hsin-Yu Chen and Adam Miller for insightful commentary on the manuscript. We are also very grateful for the assistance of Andrew Kim and William Tong in the development of the work.

\bibliographystyle{yahapj}
\bibliography{references}

\begin{thebibliography}{}
\providecommand\natexlab[1]{#1}
\providecommand\JournalTitle[1]{#1}

\bibitem[{{Aasi} {et~al.}(2015){Aasi}, {Abbott}, {Abbott}, {Abbott},
  {Abernathy}, {Ackley}, {Adams}, {Adams}, {Addesso}, \& et~al.}]{AdvLIGO}
{Aasi}, J., {Abbott}, B.~P., {Abbott}, R., {et~al.} 2015,
  \href{http://dx.doi.org/10.1088/0264-9381/32/7/074001}{\JournalTitle{Classical
  and Quantum Gravity}, 32, 074001}

\bibitem[{{Abbott} {et~al.}(2016{\natexlab{a}}){Abbott}, {Abbott}, {Abbott},
  {Abernathy}, {Acernese}, {Ackley}, {Adams}, {Adams}, {Addesso}, {Adhikari},
  \& et~al.}]{2016ApJ...826L..13A}
{Abbott}, B.~P., {Abbott}, R., {Abbott}, T.~D., {et~al.} 2016{\natexlab{a}},
  \href{http://dx.doi.org/10.3847/2041-8205/826/1/L13}{\JournalTitle{\apjl},
  826, L13}

\bibitem[{{Abbott} {et~al.}(2016{\natexlab{b}}){Abbott}, {Abbott}, {Abbott},
  {Abernathy}, {Acernese}, {Ackley}, {Adams}, {Adams}, {Addesso}, {Adhikari},
  \& et~al.}]{2016LRR....19....1A}
---. 2016{\natexlab{b}},
  \href{http://dx.doi.org/10.1007/lrr-2016-1}{\JournalTitle{Living Reviews in
  Relativity}, 19, 1}

\bibitem[{{Abbott} {et~al.}(2016{\natexlab{c}}){Abbott}, {Abbott}, {Abbott},
  {Abernathy}, {Acernese}, {Ackley}, {Adams}, {Adams}, {Addesso}, {Adhikari},
  \& et~al.}]{2016ApJ...832L..21A}
---. 2016{\natexlab{c}},
  \href{http://dx.doi.org/10.3847/2041-8205/832/2/L21}{\JournalTitle{\apjl},
  832, L21}

\bibitem[{{Abbott} {et~al.}(2017{\natexlab{a}}){Abbott}, {Abbott}, {Abbott},
  {Acernese}, {Ackley}, {Adams}, {Adams}, {Addesso}, {Adhikari}, {Adya}, \&
  et~al.}]{2017ApJ...850L..39A}
---. 2017{\natexlab{a}},
  \href{http://dx.doi.org/10.3847/2041-8213/aa9478}{\JournalTitle{\apjl}, 850,
  L39}

\bibitem[{{Abbott} {et~al.}(2017{\natexlab{b}}){Abbott}, {Abbott}, {Abbott},
  {Abernathy}, {Ackley}, {Adams}, {Addesso}, {Adhikari}, {Adya}, {Affeldt}, \&
  et~al.}]{2017CQGra..34d4001A}
---. 2017{\natexlab{b}},
  \href{http://dx.doi.org/10.1088/1361-6382/aa51f4}{\JournalTitle{Classical and
  Quantum Gravity}, 34, 044001}

\bibitem[{{Abbott} {et~al.}(2017{\natexlab{c}}){Abbott}, {Abbott}, {Abbott},
  {Acernese}, {Ackley}, {Adams}, {Adams}, {Addesso}, {Adhikari}, {Adya},
  {et~al.}}]{2017ApJ...848L..13A}
---. 2017{\natexlab{c}},
  \href{http://dx.doi.org/10.3847/2041-8213/aa920c}{\JournalTitle{\apjl}, 848,
  L13}

\bibitem[{{Abbott} {et~al.}(2017{\natexlab{d}}){Abbott}, {Abbott}, {Abbott},
  {Acernese}, {Ackley}, {Adams}, {Adams}, {Addesso}, {Adhikari}, {Adya}, \&
  et~al.}]{2017PhRvL.119n1101A}
---. 2017{\natexlab{d}},
  \href{http://dx.doi.org/10.1103/PhysRevLett.119.141101}{\JournalTitle{Physical
  Review Letters}, 119, 141101}

\bibitem[{{Abbott} {et~al.}(2017{\natexlab{e}}){Abbott}, {Abbott}, {Abbott},
  {Acernese}, {Ackley}, {Adams}, {Adams}, {Addesso}, {Adhikari}, {Adya}, \&
  et~al.}]{2017PhRvL.119p1101A}
---. 2017{\natexlab{e}},
  \href{http://dx.doi.org/10.1103/PhysRevLett.119.161101}{\JournalTitle{Physical
  Review Letters}, 119, 161101}

\bibitem[{{Abbott} {et~al.}(2017{\natexlab{f}}){Abbott}, {Abbott}, {Abbott},
  {Acernese}, {Ackley}, {Adams}, {Adams}, {Addesso}, {Adhikari}, {Adya}, \&
  et~al.}]{2017ApJ...848L..12A}
---. 2017{\natexlab{f}},
  \href{http://dx.doi.org/10.3847/2041-8213/aa91c9}{\JournalTitle{\apjl}, 848,
  L12}

\bibitem[{{Acernese} {et~al.}(2015){Acernese}, {Agathos}, {Agatsuma}, {Aisa},
  {Allemandou}, {Allocca}, {Amarni}, {Astone}, {Balestri}, {Ballardin}, \&
  et~al.}]{2015CQGra..32b4001A}
{Acernese}, F., {Agathos}, M., {Agatsuma}, K., {et~al.} 2015,
  \href{http://dx.doi.org/10.1088/0264-9381/32/2/024001}{\JournalTitle{Classical
  and Quantum Gravity}, 32, 024001}

\bibitem[{{Aso} {et~al.}(2013){Aso}, {Michimura}, {Somiya}, {Ando}, {Miyakawa},
  {Sekiguchi}, {Tatsumi}, \& {Yamamoto}}]{2013PhRvD..88d3007A}
{Aso}, Y., {Michimura}, Y., {Somiya}, K., {et~al.} 2013,
  \href{http://dx.doi.org/10.1103/PhysRevD.88.043007}{\JournalTitle{\prd}, 88,
  043007}

\bibitem[{Bardeen {et~al.}(1972)Bardeen, Press, \& Teukolsky}]{rotbh}
Bardeen, J.~M., Press, W.~H., \& Teukolsky, S.~A. 1972, \JournalTitle{The
  Astrophysical Journal}, 178, 347

\bibitem[{{Bellm}(2016)}]{2016PASP..128h4501B}
{Bellm}, E.~C. 2016,
  \href{http://dx.doi.org/10.1088/1538-3873/128/966/084501}{\JournalTitle{\pasp},
  128, 084501}

\bibitem[{{Berry} {et~al.}(2015){Berry}, {Mandel}, {Middleton}, {Singer},
  {Urban}, {Vecchio}, {Vitale}, {Cannon}, {Farr}, {Farr}, {Graff}, {Hanna},
  {Haster}, {Mohapatra}, {Pankow}, {Price}, {Sidery}, \&
  {Veitch}}]{2015ApJ...804..114B}
{Berry}, C.~P.~L., {Mandel}, I., {Middleton}, H., {et~al.} 2015,
  \href{http://dx.doi.org/10.1088/0004-637X/804/2/114}{\JournalTitle{\apj},
  804, 114}

\bibitem[{{Chen} \& {Holz}(2016)}]{2016arXiv161201471C}
{Chen}, H.-Y., \& {Holz}, D.~E. 2016, \JournalTitle{ArXiv e-prints},
  \href{http://arxiv.org/abs/1612.01471}{{\sffamily arXiv:1612.01471
  [astro-ph.HE]}}

\bibitem[{{Chen} {et~al.}(2017){Chen}, {Holz}, {Miller}, {Evans}, {Vitale}, \&
  {Creighton}}]{2017arXiv170908079C}
{Chen}, H.-Y., {Holz}, D.~E., {Miller}, J., {et~al.} 2017, \JournalTitle{ArXiv
  e-prints}, \href{http://arxiv.org/abs/1709.08079}{{\sffamily
  arXiv:1709.08079}}

\bibitem[{Collaboration {et~al.}(2017)Collaboration, Virgo, Collaboration,
  Collaboration, the DES, Collaboration, Collaboration, Collaboration, \&
  Collaboration}]{LVCHubble2017}
Collaboration, T. L. S.~C., Virgo, T., Collaboration, T.~M., {et~al.} 2017,
  \href{http://dx.doi.org/10.1038/nature24471}{\JournalTitle{Nature}, 551, 85},
  letter

\bibitem[{{Corral-Santana} {et~al.}(2016){Corral-Santana}, {Casares},
  {Mu{\~n}oz-Darias}, {Bauer}, {Mart{\'{\i}}nez-Pais}, \&
  {Russell}}]{2016A&A...587A..61C}
{Corral-Santana}, J.~M., {Casares}, J., {Mu{\~n}oz-Darias}, T., {et~al.} 2016,
  \href{http://dx.doi.org/10.1051/0004-6361/201527130}{\JournalTitle{\aap},
  587, A61}

\bibitem[{Cowperthwaite \& Berger(2015)}]{0004-637X-814-1-25}
Cowperthwaite, P.~S., \& Berger, E. 2015,
  \href{http://stacks.iop.org/0004-637X/814/i=1/a=25}{\JournalTitle{The
  Astrophysical Journal}, 814, 25}

\bibitem[{{Fairhurst}(2011{\natexlab{a}})}]{2011CQGra..28j5021F}
{Fairhurst}, S. 2011{\natexlab{a}},
  \href{http://dx.doi.org/10.1088/0264-9381/28/10/105021}{\JournalTitle{Classical
  and Quantum Gravity}, 28, 105021}

\bibitem[{{Fairhurst}(2011{\natexlab{b}})}]{2011NJPh...13f9602F}
---. 2011{\natexlab{b}},
  \href{http://dx.doi.org/10.1088/1367-2630/13/6/069602}{\JournalTitle{New
  Journal of Physics}, 13, 069602}

\bibitem[{{Farr} {et~al.}(2016){Farr}, {Berry}, {Farr}, {Haster}, {Middleton},
  {Cannon}, {Graff}, {Hanna}, {Mandel}, {Pankow}, {Price}, {Sidery}, {Singer},
  {Urban}, {Vecchio}, {Veitch}, \& {Vitale}}]{2016ApJ...825..116F}
{Farr}, B., {Berry}, C.~P.~L., {Farr}, W.~M., {et~al.} 2016,
  \href{http://dx.doi.org/10.3847/0004-637X/825/2/116}{\JournalTitle{\apj},
  825, 116}

\bibitem[{Farr {et~al.}(2011)Farr, Sravan, Cantrell, Kreidberg, Bailyn, Mandel,
  \& Kalogera}]{0004-637X-741-2-103}
Farr, W.~M., Sravan, N., Cantrell, A., {et~al.} 2011,
  \href{http://stacks.iop.org/0004-637X/741/i=2/a=103}{\JournalTitle{The
  Astrophysical Journal}, 741, 103}

\bibitem[{Flaugher {et~al.}(2015)}]{Flaugher:2015pxc}
Flaugher, B., {et~al.} 2015,
  \href{http://dx.doi.org/10.1088/0004-6256/150/5/150}{\JournalTitle{Astron.
  J.}, 150, 150}

\bibitem[{Fragos \& McClintock(2015)}]{0004-637X-800-1-17}
Fragos, T., \& McClintock, J.~E. 2015,
  \href{http://stacks.iop.org/0004-637X/800/i=1/a=17}{\JournalTitle{The
  Astrophysical Journal}, 800, 17}

\bibitem[{{Freire} {et~al.}(2008){Freire}, {Ransom}, {B{\'e}gin}, {Stairs},
  {Hessels}, {Frey}, \& {Camilo}}]{2008ApJ...675..670F}
{Freire}, P.~C.~C., {Ransom}, S.~M., {B{\'e}gin}, S., {et~al.} 2008,
  \href{http://dx.doi.org/10.1086/526338}{\JournalTitle{\apj}, 675, 670}

\bibitem[{{Gaebel} \& {Veitch}(2017)}]{2017arXiv170308988G}
{Gaebel}, S.~M., \& {Veitch}, J. 2017, \JournalTitle{ArXiv e-prints},
  \href{http://arxiv.org/abs/1703.08988}{{\sffamily arXiv:1703.08988
  [astro-ph.IM]}}

\bibitem[{{Hannam} {et~al.}(2014){Hannam}, {Schmidt}, {Boh{\'e}}, {Haegel},
  {Husa}, {Ohme}, {Pratten}, \& {P{\"u}rrer}}]{2014PhRvL.113o1101H}
{Hannam}, M., {Schmidt}, P., {Boh{\'e}}, A., {et~al.} 2014,
  \href{http://dx.doi.org/10.1103/PhysRevLett.113.151101}{\JournalTitle{Physical
  Review Letters}, 113, 151101}

\bibitem[{{Hu} {et~al.}(2015){Hu}, {Raffai}, {Gond{\'a}n}, {Heng},
  {Kelecs{\'e}nyi}, {Hendry}, {M{\'a}rka}, \&
  {M{\'a}rka}}]{2015CQGra..32j5010H}
{Hu}, Y.-M., {Raffai}, P., {Gond{\'a}n}, L., {et~al.} 2015,
  \href{http://dx.doi.org/10.1088/0264-9381/32/10/105010}{\JournalTitle{Classical
  and Quantum Gravity}, 32, 105010}

\bibitem[{{IndIGO Collaboration}(2011)}]{ligoindia}
{IndIGO Collaboration}. 2011, LIGO-India, Proposal of the Consortium for Indian
  Initiative in Gravitational-wave Observations (IndIGO), Tech. Rep.
  LIGO-M1100296-v2, IndIGO,
  https://dcc.ligo.org/cgi-bin/DocDB/ShowDocument?docid=75988

\bibitem[{{Ivezic} {et~al.}(2008){Ivezic}, {Tyson}, {Abel}, {Acosta},
  {Allsman}, {AlSayyad}, {Anderson}, {Andrew}, {Angel}, {Angeli}, {Ansari},
  {Antilogus}, {Arndt}, {Astier}, {Aubourg}, {Axelrod}, {Bard}, {Barr},
  {Barrau}, {Bartlett}, {Bauman}, {Beaumont}, {Becker}, {Becla}, {Beldica},
  {Bellavia}, {Blanc}, {Blandford}, {Bloom}, {Bogart}, {Borne}, {Bosch},
  {Boutigny}, {Brandt}, {Brown}, {Bullock}, {Burchat}, {Burke}, {Cagnoli},
  {Calabrese}, {Chandrasekharan}, {Chesley}, {Cheu}, {Chiang}, {Claver},
  {Connolly}, {Cook}, {Cooray}, {Covey}, {Cribbs}, {Cui}, {Cutri}, {Daubard},
  {Daues}, {Delgado}, {Digel}, {Doherty}, {Dubois}, {Dubois-Felsmann},
  {Durech}, {Eracleous}, {Ferguson}, {Frank}, {Freemon}, {Gangler}, {Gawiser},
  {Geary}, {Gee}, {Geha}, {Gibson}, {Gilmore}, {Glanzman}, {Goodenow},
  {Gressler}, {Gris}, {Guyonnet}, {Hascall}, {Haupt}, {Hernandez}, {Hogan},
  {Huang}, {Huffer}, {Innes}, {Jacoby}, {Jain}, {Jee}, {Jernigan},
  {Jevremovic}, {Johns}, {Jones}, {Juramy-Gilles}, {Juric}, {Kahn}, {Kalirai},
  {Kallivayalil}, {Kalmbach}, {Kantor}, {Kasliwal}, {Kessler}, {Kirkby},
  {Knox}, {Kotov}, {Krabbendam}, {Krughoff}, {Kubanek}, {Kuczewski},
  {Kulkarni}, {Lambert}, {Le Guillou}, {Levine}, {Liang}, {Lim}, {Lintott},
  {Lupton}, {Mahabal}, {Marshall}, {Marshall}, {May}, {McKercher}, {Migliore},
  {Miller}, {Mills}, {Monet}, {Moniez}, {Neill}, {Nief}, {Nomerotski},
  {Nordby}, {O'Connor}, {Oliver}, {Olivier}, {Olsen}, {Ortiz}, {Owen}, {Pain},
  {Peterson}, {Petry}, {Pierfederici}, {Pietrowicz}, {Pike}, {Pinto}, {Plante},
  {Plate}, {Price}, {Prouza}, {Radeka}, {Rajagopal}, {Rasmussen}, {Regnault},
  {Ridgway}, {Ritz}, {Rosing}, {Roucelle}, {Rumore}, {Russo}, {Saha},
  {Sassolas}, {Schalk}, {Schindler}, {Schneider}, {Schumacher}, {Sebag},
  {Sembroski}, {Seppala}, {Shipsey}, {Silvestri}, {Smith}, {Smith}, {Strauss},
  {Stubbs}, {Sweeney}, {Szalay}, {Takacs}, {Thaler}, {Van Berg}, {Vanden Berk},
  {Vetter}, {Virieux}, {Xin}, {Walkowicz}, {Walter}, {Wang}, {Warner},
  {Willman}, {Wittman}, {Wolff}, {Wood-Vasey}, {Yoachim}, {Zhan}, \& {for the
  LSST Collaboration}}]{2008arXiv0805.2366I}
{Ivezic}, Z., {Tyson}, J.~A., {Abel}, B., {et~al.} 2008, \JournalTitle{ArXiv
  e-prints}, \href{http://arxiv.org/abs/0805.2366}{{\sffamily arXiv:0805.2366}}

\bibitem[{{Kramer} \& {Wex}(2009)}]{2009CQGra..26g3001K}
{Kramer}, M., \& {Wex}, N. 2009,
  \href{http://dx.doi.org/10.1088/0264-9381/26/7/073001}{\JournalTitle{Classical
  and Quantum Gravity}, 26, 073001}

\bibitem[{{LIGO Scientific Collaboration} {et~al.}(2015){LIGO Scientific
  Collaboration}, {Aasi}, {Abbott}, {Abbott}, {Abbott}, {Abernathy}, {Ackley},
  {Adams}, {Adams}, {Addesso}, \& et~al.}]{2015CQGra..32g4001L}
{LIGO Scientific Collaboration}, {Aasi}, J., {Abbott}, B.~P., {et~al.} 2015,
  \href{http://dx.doi.org/10.1088/0264-9381/32/7/074001}{\JournalTitle{Classical
  and Quantum Gravity}, 32, 074001}

\bibitem[{Messick {et~al.}(2017)Messick, Blackburn, Brady, Brockill, Cannon,
  Cariou, Caudill, Chamberlin, Creighton, Everett, Hanna, Keppel, Lang, Li,
  Meacher, Nielsen, Pankow, Privitera, Qi, Sachdev, Sadeghian, Singer, Thomas,
  Wade, Wade, Weinstein, \& Wiesner}]{PhysRevD.95.042001}
Messick, C., Blackburn, K., Brady, P., {et~al.} 2017,
  \href{http://dx.doi.org/10.1103/PhysRevD.95.042001}{\JournalTitle{Phys. Rev.
  D}, 95, 042001}

\bibitem[{Metzger \& Berger(2012)}]{0004-637X-746-1-48}
Metzger, B.~D., \& Berger, E. 2012,
  \href{http://stacks.iop.org/0004-637X/746/i=1/a=48}{\JournalTitle{The
  Astrophysical Journal}, 746, 48}

\bibitem[{{Nissanke} {et~al.}(2013){Nissanke}, {Kasliwal}, \&
  {Georgieva}}]{2013ApJ...767..124N}
{Nissanke}, S., {Kasliwal}, M., \& {Georgieva}, A. 2013,
  \href{http://dx.doi.org/10.1088/0004-637X/767/2/124}{\JournalTitle{\apj},
  767, 124}

\bibitem[{{Nissanke} {et~al.}(2011){Nissanke}, {Sievers}, {Dalal}, \&
  {Holz}}]{2011ApJ...739...99N}
{Nissanke}, S., {Sievers}, J., {Dalal}, N., \& {Holz}, D. 2011,
  \href{http://dx.doi.org/10.1088/0004-637X/739/2/99}{\JournalTitle{\apj}, 739,
  99}

\bibitem[{{Pankow} {et~al.}(2017){Pankow}, {Sampson}, {Perri}, {Chase},
  {Coughlin}, {Zevin}, \& {Kalogera}}]{2017ApJ...834..154P}
{Pankow}, C., {Sampson}, L., {Perri}, L., {et~al.} 2017,
  \href{http://dx.doi.org/10.3847/1538-4357/834/2/154}{\JournalTitle{\apj},
  834, 154}

\bibitem[{Pannarale \& Ohme(2014)}]{2041-8205-791-1-L7}
Pannarale, F., \& Ohme, F. 2014,
  \href{http://stacks.iop.org/2041-8205/791/i=1/a=L7}{\JournalTitle{The
  Astrophysical Journal Letters}, 791, L7}

\bibitem[{{Patricelli} {et~al.}(2016){Patricelli}, {Razzano}, {Cella},
  {Fidecaro}, {Pian}, {Branchesi}, \& {Stamerra}}]{2016JCAP...11..056P}
{Patricelli}, B., {Razzano}, M., {Cella}, G., {et~al.} 2016,
  \href{http://dx.doi.org/10.1088/1475-7516/2016/11/056}{\JournalTitle{\jcap},
  11, 056}

\bibitem[{Raymond {et~al.}(2009)Raymond, van~der Sluys, Mandel, Kalogera,
  Röver, \& Christensen}]{0264-9381-26-11-114007}
Raymond, V., van~der Sluys, M.~V., Mandel, I., {et~al.} 2009,
  \href{http://stacks.iop.org/0264-9381/26/i=11/a=114007}{\JournalTitle{Classical
  and Quantum Gravity}, 26, 114007}

\bibitem[{{Reisswig} {et~al.}(2009){Reisswig}, {Husa}, {Rezzolla}, {Dorband},
  {Pollney}, \& {Seiler}}]{2009PhRvD..80l4026R}
{Reisswig}, C., {Husa}, S., {Rezzolla}, L., {et~al.} 2009,
  \href{http://dx.doi.org/10.1103/PhysRevD.80.124026}{\JournalTitle{\prd}, 80,
  124026}

\bibitem[{{Rodriguez} {et~al.}(2014){Rodriguez}, {Farr}, {Raymond}, {Farr},
  {Littenberg}, {Fazi}, \& {Kalogera}}]{2014ApJ...784..119R}
{Rodriguez}, C.~L., {Farr}, B., {Raymond}, V., {et~al.} 2014,
  \href{http://dx.doi.org/10.1088/0004-637X/784/2/119}{\JournalTitle{\apj},
  784, 119}

\bibitem[{{Searle} {et~al.}(2006){Searle}, {Scott}, {McClelland}, \&
  {Finn}}]{2006PhRvD..73l4014S}
{Searle}, A.~C., {Scott}, S.~M., {McClelland}, D.~E., \& {Finn}, L.~S. 2006,
  \href{http://dx.doi.org/10.1103/PhysRevD.73.124014}{\JournalTitle{\prd}, 73,
  124014}

\bibitem[{{Singer} \& {Price}(2016)}]{2016PhRvD..93b4013S}
{Singer}, L.~P., \& {Price}, L.~R. 2016,
  \href{http://dx.doi.org/10.1103/PhysRevD.93.024013}{\JournalTitle{\prd}, 93,
  024013}

\bibitem[{{Singer} {et~al.}(2014){Singer}, {Price}, {Farr}, {Urban}, {Pankow},
  {Vitale}, {Veitch}, {Farr}, {Hanna}, {Cannon}, {Downes}, {Graff}, {Haster},
  {Mandel}, {Sidery}, \& {Vecchio}}]{2014ApJ...795..105S}
{Singer}, L.~P., {Price}, L.~R., {Farr}, B., {et~al.} 2014,
  \href{http://dx.doi.org/10.1088/0004-637X/795/2/105}{\JournalTitle{\apj},
  795, 105}

\bibitem[{{Smartt} {et~al.}(2017){Smartt}, {Chen}, {Jerkstrand}, {Coughlin}, \&
  et~al.}]{2017Natur.551...75S}
{Smartt}, S.~J., {Chen}, T.-W., {Jerkstrand}, A., {Coughlin}, M., \& et~al.
  2017, \href{http://dx.doi.org/10.1038/nature24303}{\JournalTitle{\nat}, 551,
  75}

\bibitem[{Vallisneri(2008)}]{PhysRevD.77.042001}
Vallisneri, M. 2008,
  \href{http://dx.doi.org/10.1103/PhysRevD.77.042001}{\JournalTitle{Phys. Rev.
  D}, 77, 042001}

\bibitem[{{van der Sluys} {et~al.}(2008){van der Sluys}, {R{\"o}ver},
  {Stroeer}, {Raymond}, {Mandel}, {Christensen}, {Kalogera}, {Meyer}, \&
  {Vecchio}}]{2008ApJ...688L..61V}
{van der Sluys}, M.~V., {R{\"o}ver}, C., {Stroeer}, A., {et~al.} 2008,
  \href{http://dx.doi.org/10.1086/595279}{\JournalTitle{\apjl}, 688, L61}

\bibitem[{{Veitch} {et~al.}(2012){Veitch}, {Mandel}, {Aylott}, {Farr},
  {Raymond}, {Rodriguez}, {van der Sluys}, {Kalogera}, \&
  {Vecchio}}]{2012PhRvD..85j4045V}
{Veitch}, J., {Mandel}, I., {Aylott}, B., {et~al.} 2012,
  \href{http://dx.doi.org/10.1103/PhysRevD.85.104045}{\JournalTitle{\prd}, 85,
  104045}

\bibitem[{{Veitch} {et~al.}(2015){Veitch}, {Raymond}, {Farr}, {Farr}, {Graff},
  {Vitale}, {Aylott}, {Blackburn}, {Christensen}, {Coughlin}, {Del Pozzo},
  {Feroz}, {Gair}, {Haster}, {Kalogera}, {Littenberg}, {Mandel},
  {O'Shaughnessy}, {Pitkin}, {Rodriguez}, {R{\"o}ver}, {Sidery}, {Smith}, {Van
  Der Sluys}, {Vecchio}, {Vousden}, \& {Wade}}]{2015PhRvD..91d2003V}
{Veitch}, J., {Raymond}, V., {Farr}, B., {et~al.} 2015,
  \href{http://dx.doi.org/10.1103/PhysRevD.91.042003}{\JournalTitle{\prd}, 91,
  042003}

\bibitem[{{Vitale} \& {Zanolin}(2011)}]{2011PhRvD..84j4020V}
{Vitale}, S., \& {Zanolin}, M. 2011,
  \href{http://dx.doi.org/10.1103/PhysRevD.84.104020}{\JournalTitle{\prd}, 84,
  104020}

\bibitem[{{Wen} \& {Chen}(2010)}]{2010PhRvD..81h2001W}
{Wen}, L., \& {Chen}, Y. 2010,
  \href{http://dx.doi.org/10.1103/PhysRevD.81.082001}{\JournalTitle{\prd}, 81,
  082001}

\end{thebibliography}

\end{document}